\documentclass{pasj00}
\draft

\usepackage{color}
\usepackage{url}
\usepackage{bm}
\usepackage{multirow}

\begin{document}
\SetRunningHead{Author(s) in page-head}{Running Head}

\title{Comparison between Hinode/SOT and SDO/HMI, AIA Data for the Study of the Solar Flare Trigger Process}

\author{Yumi \textsc{Bamba}, Kanya \textsc{Kusano} \altaffilmark{1} and Shinsuke \textsc{Imada}}
\affil{Solar-Terrestrial Environment Laboratory, Nagoya University,
Furo-cho, Chikusa-ku, Nagoya, Aichi 464-8601, Japan}
\altaffiltext{1}{Japan Agency for Marine-Earth Science and Technology (JAMSTEC),
Kanazawa-ku, Yokohama, Kanagawa, 2360001, Japan}
\email{y-bamba@stelab.nagoyau.ac.jp}
\email{kusano@nagoya-u.jp}
\email{shinimada@stelab.nagoya-u.ac.jp}
\and
\author{Yusuke \textsc{Iida}}
\affil{Institute of Space and Astronautical Science, Japan Aerospace Exploration Agency,
Chuo-ku, Sagamihara, Kanagawa 252-5210, Japan}
\email{iida@solar.isas.jaxa.jp}

%


\maketitle

\begin{abstract}
Understanding the mechanism that produces solar flares is important not only from the scientific point of view but also for improving space weather predictability. There are numerous observational and computational studies, which attempted to reveal the onset mechanism of solar flares. However, the underlying mechanism of flare onset remains elusive. To elucidate the flare trigger mechanism, we have analyzed several flare events which were observed by Hinode/Solar Optical Telescope (SOT), in our previous study. Because of the limitation of SOT field of view, however, only four events in the Hinode data sets have been utilizable. Therefore, increasing the number of events is required for evaluating the flare trigger models. 

We investigated the applicability of data obtained by the Solar Dynamics Observatory (SDO) to increase the data sample for a statistical analysis of the flare trigger process. SDO regularly observes the full disk of the sun and all flares although its spatial resolution is lower than that of Hinode. We investigated the M6.6 flare which occurred on 13 February 2011 and compared the analyzed data of SDO with the results of our previous study using Hinode/SOT data. Filter and vector magnetograms obtained by the Helioseismic and Magnetic Imager (HMI) and filtergrams from the Atmospheric Imaging Assembly (AIA) 1600{\AA} were employed. From the comparison of small-scale magnetic configurations and chromospheric emission prior to the flare onset, we confirmed that the trigger region is detectable with the SDO data. We also measured the magnetic shear angles of the active region and the azimuth and strength of the flare-trigger field. The results were consistent with our previous study. We concluded that statistical studies of the flare trigger process are feasible with SDO as well as Hinode data. We also investigated the temporal evolution of the magnetic field before the flare onset with SDO.
\end{abstract}

\section{Introduction} \label{sec:intro}

Solar flares are explosive phenomena which release magnetic energy stored in the solar corona as thermal and kinetic energy of the plasma. Because solar flares sometimes cause interplanetary disturbances and impact our terrestrial environment, it is important to understand the mechanism of solar flares also from the practical point of view. There are numerous observational studies and simulations which attempted to reveal the onset mechanism of solar flares (e.g., \cite{heyvaerts77}, \cite{antiochos99}, \cite{chen00}, \cite{moore01}). From these previous studies, it is widely accepted that magnetic reconnection plays important roles in the triggering process of solar flares (e.g., \cite{yokoyama98}). However, the underlying mechanism of flare onset remains elusive, and the predictability of flare occurrence remains limited.

Recently, \citet{bamba13} analyzed several flare events observed by the Solar Optical Telescope (SOT; \cite{tsuneta08, suematsu08, ichimoto08, shimizu07}) onboard the Hinode satellite (\cite{kosugi07}), in order to elucidate the flare trigger mechanism. They investigated the spatio-temporal correlation between the detailed magnetic field structure and the chromospheric pre-flare emission at the central part of flaring regions for several hours prior to the onset of flares. They found that the magnetic shear angle in the flaring regions exceeded 70 degrees, as well as that characteristic magnetic disturbances developed at the centers of flaring regions in the pre-flare phase. The observed signatures strongly support the flare trigger mechanism presented by \citet{kusano12}, who proposed that solar flares can be triggered by the interaction between the sheared arcade and one of two types of small magnetic structures: ``Opposite-Polarity (OP)'' and ``Reversed-Shear (RS)''.

Because of the limitation of the SOT field of view (FOV), only four data sets were utilizable in \citet{bamba13}. Increasing the number of events is required for more precisely verifying the trigger model proposed by \citet{kusano12}. Solar Dynamics Observatory (SDO; \cite{pesnell12}) observes the full disk of the sun and all flares, although its spatial resolution is lower than that of Hinode/SOT. Therefore, in this paper, we evaluated whether the data sets obtained by SDO are usable for the detailed analysis of small magnetic structure and chromospheric emission in the pre-flare phase. We choose the M6.6 flare which occurred on 13 February 2011 and was observed both by Hinode and SDO, as a sample event. \citet{bamba13} already analyzed this event using Hinode data, and here we checked whether we can obtain their same conclusions using only the SDO data. If we can obtain the same conclusion with \citet{bamba13}, it would indicate that we will be able to extend the analysis methods of \citet{bamba13} to events seen by SDO. This would substantially expand the number of events available for study with these methods. In addition, we also extend our vector magnetic field analysis of the 13 February 2011 flare to earlier phase, which could not be analyzed in the previous study with the Hinode data alone.

This paper is organized as follows. The sampled data sets and their analysis method are described in Section \ref{sec:data}. We compared the results with Hinode and SDO, and the possibility of new analysis with the SDO data sets is presented in Section \ref{sec:results}. We discuss the interpretation of the results in Section \ref{sec:discussion}. Finally, we summarize our study in Section \ref{sec:summary}.

\section{Data and Analysis} \label{sec:data}

\subsection{Data Description} \label{sec:data1}

The active region (AR) NOAA 11158 appeared in the southern hemisphere near solar disc center on 11 February 2011. Although a multitude of flares including one X-class flare and five M-class flares were produced in the active region, we focused on the M6.6 flare which occurred on February 13 in this paper. The onset time of the M6.6 flare was 17:28 UT according to the soft X-ray data obtained by GOES (Geostationary Operational Environment) 1-8{\AA}. The M6.6 flare was already analyzed in \citet{bamba13} and \citet{toriumi13} with Hinode/SOT data, and they concluded that the flare was triggered by RS-type small magnetic disturbance that appeared on the major polarity inversion line (PIL) of the central bipole of the active region. The Helioseismic and Magnetic Imager (HMI; \cite{schou}) and the Atmospheric Imaging Assembly (AIA; \cite{lemen}) onboard the SDO satellite also observed the active region before and after the M6.6 flare.

In the previous study \citep{bamba13}, they employed Ca II H line (3968{\AA}) filtergrams and Stokes-V/I images in the Na D1 line (5896{\AA}) obtained by Hinode/SOT. The Ca II H line is sensitive to emission from the upper chromosphere. The M6.6 flare was observed in these wavelengths and at a cadence of about five minutes. Unfortunately, there is a data gap from 13:15 UT to 15:00 UT. The FOV of SOT was limited to $328" \times 164"$ for the Narrowband Filter Imager (NFI) and $218" \times 109"$ for the Broadband Filter Imager (BFI), as mentioned in Section \ref{sec:intro}. The spatial resolutions of the NFI and BFI are about $0.3"$ and $0.2"$, respectively. They also used the vector magnetograms obtained by the Spectro-Polarimeter (SP) of SOT \citep{lites13}. The SP observed the full-polarization states (Stokes-I, Q, U, and V) of two magnetically sensitive Fe lines at $6301.5$ and $6302.5${\AA} at 16:00:04-16:32:25 UT on February 13. The FOV of the vector magnetograms was $151" \times 162"$.

SDO observed the sun with a full disk FOV ($2000" \times 2000"$) at 45 seconds cadence before and throughout the flare. The spatial resolution is $1"$ for HMI and $1.5"$ for AIA. In this study, we choose AIA continuum and C IV line (1600{\AA}) filtergrams and HMI filter magnetograms in the Fe I line (6173{\AA}), which are sensitive to emission from the transition region and upper chromosphere, and photospheric magnetic field, respectively. We also used HMI vector magnetograms in the Fe I line (6173{\AA}) at 12 minutes cadence, in order to measure the magnetic shear angle of the active region. These vector magnetograms were extracted from SHARP (Spaceweather HMI Active Region Patche) data series in which the $180^{\circ}$ ambiguity was resolved and the magnetic field vector has been remapped to a Lambert Cylindrical Equal-Area projection. Table \ref{tab:table1} indicates the summary of the filtergraph instrumentation and each parameter of Hinode/SOT and SDO/HMI, AIA.

\begin{table}
  \caption{Summary of the filtergraph instrumentation and each parameter of Hinode/SOT and SDO/HMI, AIA}\label{tab:table1}
  \begin{center}
    \begin{tabular}{|c||c|c|c|c|}
      \hline
      \multirow{2}{*}{instrument} & \multicolumn{2}{c|}{Hinode } & \multicolumn{2}{c|}{SDO} \\ \cline{2-5}
   & SOT/NFI & SOT/BFI & HMI & AIA \\ \hline \hline
      FOV & $328" \times 164"$ & $218" \times 109"$ & $2000" \times 2000"$ & $2000" \times 2000"$ \\ \hline
  spatial resolution & $0.3"$ & $0.2"$ & $1.0"$ & $1.5"$ \\ \hline
  wavelength & 5896 {\AA} & 3968 {\AA} & 6173 {\AA} & 1600 {\AA} \\ \hline
  primary ion(s) & Na & Ca II H & Fe I & C IV + continuum \\ \hline
  \multirow{3}{*}{object(s)} & upper photospheric  & lower chromospheric  & lower photospheric & transition region \\
  & lower chromospheric & emission & magnetic field & upper photospheric \\
  & magnetic field & & & emission \\
  \hline
    \end{tabular}
  \end{center}
\end{table}

\subsection{Method of Analysis} \label{sec:data2}

\begin{figure}
 \begin{center}
  \includegraphics[width=10cm,clip]{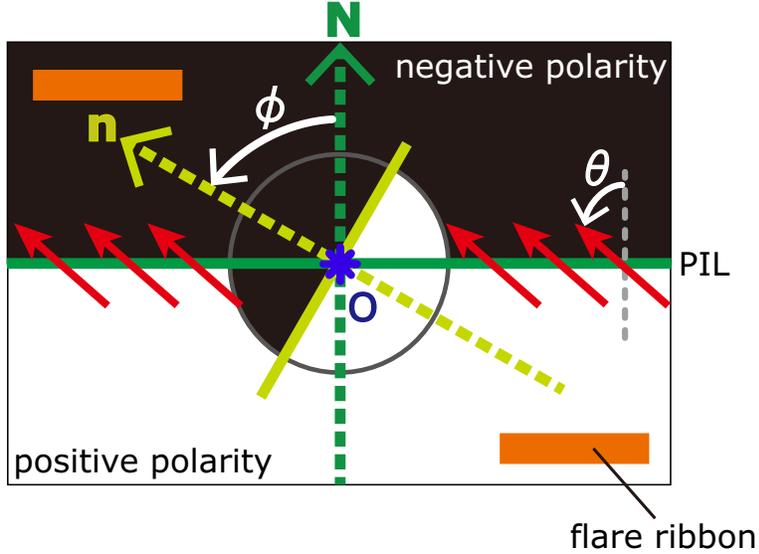} 
 \end{center}
\caption{
Illustration of the shear angle $\theta$ and the azimuth angle $\phi$.
White/black corresponds to the positive/negative polarity of the LOS magnetic field and the orange lines represent the flare ribbon. Point $\bm{O}$ is the center of the trigger region. The green-colored solid line indicates the global PIL of the active region and the vector $\bm{N}$ (green-colored dashed arrow) is normal to the global PIL. The yellow-green solid line indicates the local PIL of the flare trigger region and $\bm{n}$ (yellow-green dashed arrow) is the normal vector of the local PIL. The azimuthal angle $\phi$ is defined as the angle between vectors $\bm{N}$ and $\bm{n}$. The shear angle $\theta$ is measured as the mean angle between vector $\bm{N}$ and the transvers magnetic field vectors (red-colored arrows).
}\label{fig:fig1}
\end{figure}

In this study, we examined the applicability of the analysis, which was developed in the previous study, to the SDO data sets. The analysis method proposed by \citet{bamba13} is summarized as follows:
\begin{enumerate}
\item Stokes-V/I images and strong Ca-line emissions are superimposed to detect the ``flare trigger region'', which should be located at the center of the sheared ribbons, through the following procedure:
 \begin{itemize}
 \item Filtergram images for Ca and Stokes-V/I, which were taken almost simultaneously, were selected and resampled to the same image size.
 \item The positions of the Ca-line image and Stokes-V/I image were aligned by maximizing the cross-correlation between the two images.
 \item The Ca-line image is superimposed onto the Stokes-V/I image, and the PILs (lines of zero Stokes-V/I) and some contours for strong Ca-line emission were over-plotted on the Stokes-V/I image.
 \end{itemize}
\item The shear angle $\theta$ and the azimuth angle $\phi$, which are defined in Figure \ref{fig:fig1}, are measured by the following procedures:
\begin{itemize}
\item Point $\bm{O}$ (blue asterisk) is defined as the nearest point to the last strong Ca-line emission before the flare onset on the PIL of the smoothed magnetic field component (green-colored solid line), which is produced by low-pass filtering viz.
\begin{equation}
<B_{LOS}> = \sum_{|k_x| < k_{x0}, |k_y| < k_{y0}} \tilde{B}_{LOS}(\bm{k}) e^{i\bm{k} \cdot \bm{r}},
\label{equation:lowpass}
\end{equation}
where $\tilde{B}_{LOS}(\bm{k})$ is the complex Fourier component of mode $\bm{k}$. The critical scale $2\pi/k_{x0} = 2\pi/k_{y0} = 2.4\times10^{7}$ cm.
\item Vector $\bm{N}$ (green-colored dashed arrow) is defined as a vector normal to the smoothed PIL at point $\bm{O}$ on the data of the last strong Ca-line emission before the flare onset.
\item The angle between the transverse field (red arrows) averaged over the flaring region and the vector $\bm{N}$ is measured as the shear angle $\theta$.
\item Vector $\bm{n}$ (yellow-green dashed line) is determined as the vector normal to the non-smoothed PIL (yellow-green solid line) in the pre-flaring region, which corresponds to the region where a strong Ca-line emission was observed on the non-smoothed PIL. Here, we use the strong Ca-line emission as a proxy of the flare trigger region, because internal reconnection in the chromosphere, which may produce the Ca-line emission, can play a role for triggering flares, as suggested by \citet{kusano12}.
\item The orientation $\phi$ is calculated by averaging the angle between vectors $\bm{N}$ and $\bm{n}$ over the flare trigger region corresponding to the strong Ca-line emission. The range of the angle (maximum to minimum) was adopted as the error range of $\phi$.
\end{itemize}
\end{enumerate}

We applied this analysis method to HMI filter magnetograms and AIA 1600{\AA} images. Each filter magnetogram was processed by the {\it aia\_prep} procedure in order to sample them to the same size as the AIA images. Then we defined the flare-trigger region for the images at 17:25 UT, and measured the shear angle $\theta$ and the azimuth angle $\phi$ from 11:00 UT to 18:00 UT. Here, we skipped the data in which the AIA 1600{\AA} emission is lower than $10^{3.2}$ DN.

\section{Results} \label{sec:results}

\subsection{Hinode Data Analysis Results} \label{sec:results1}

Here we review the results of the previous study with Hinode before showing the new results. Figure \ref{fig:fig2} (a-c) shows the superposed images of Hinode/SOT. The background grayscale images correspond to the positive/negative polarity of the LOS magnetic field. The green lines indicate the PILs, and the strong Ca-line emission is overlaid with red contours. We overlaid the flare ribbon which was initially seen at 17:35 UT on the Stokes-V/I image at 17:30 UT in panel (c), where Stokes-V/I signals before the flare were used because it was largely affected by flare. The observations indicated that the onset of the flare occurred between 17:25 UT and 17:35 UT.
According to numerical simulations, \citet{kusano12} suggested the following features for the flare-trigger region:
\begin{enumerate}
\item The flare-trigger region locates around the center of the initial sheared ribbon.
\item Chromospheric brightenings might be continuously or intermittently observed on the PIL of the trigger region from several hours prior to the flare onset as a consequence of internal magnetic reconnection between the large scale arcade in the flaring region and the small arcade in the flare trigger region (see also \citet{brooks08}).
\item The shear angle $\theta$ and the azimuth angle $\phi$ must satisfy the condition represented in the ``flare phase diagram'' (Figure 2 in \citet{kusano12}) for triggering flares. The flare trigger regions of $\theta$ and $\phi$ can be classified as OP- or RS-type.
\end{enumerate}

We can see the initial sheared flare-ribbon in Figure \ref{fig:fig2} (c). There is a small positive wedge-like structure marked by a yellow circle at the center of the sheared ribbon. The transient brightening in the Ca-line indicated by a white arrow in panel (b) is seen on the wedge-like structure. This brightening had been observed about 4.5 hours prior to the flare onset accompanied by the formation of the wedge-like structure. We note that the precursor brightening was always seen on the west-side of the wedge-like structure on the PIL. Because the global shear of the active region is northeastward, the brightening on the west-side of the wedge-like structure satisfied the geometrical condition of the RS-type flare-trigger field. Therefore, \citet{bamba13} concluded that the wedge-like structure is the trigger region of the M6.6 flare. In addition, they measured the shear angle $\theta$ in the wedge-like structure and the azimuth angle $\phi$ in the flare trigger region. The results were $\theta = 82\pm7^{\circ}$, $\phi = 318^{\circ}$-$331^{\circ}$. Therefore, they confirmed that the trigger process of the M6.6 flare is consistent with the RS-type model of \citet{kusano12}.

Furthermore, they investigated the temporal evolution of the magnetic flux in the trigger region from 11:00 UT to 18:00 UT. Although there is a data gap from 13:15 UT to 15:00 UT as mentioned in Section \ref{sec:data1}, they found a flux increase in the flare-trigger field of 20\%-30\% before the flare onset.

\subsection{SDO Data Analysis Results} \label{sec:results2}

\begin{figure}
 \begin{center}
  \includegraphics[width=15cm,clip]{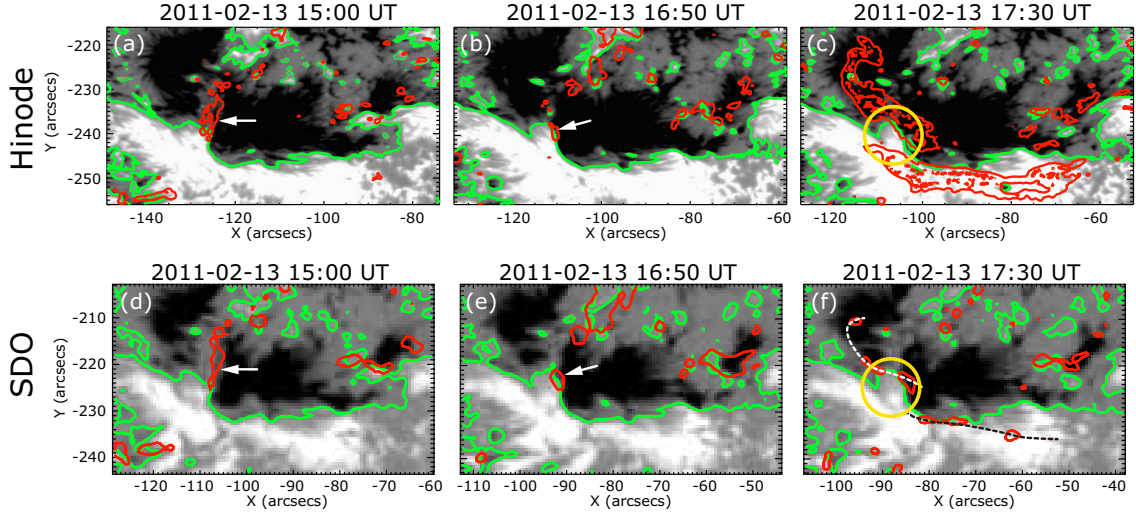} 
 \end{center}
\caption{
Temporal evolutions of magnetic structures and chromospheric emission obtained by Hinode and SDO before the M6.6 flare.
(a-c) show the observations of Hinode/SOT. The background is the Stokes-V/I image and white/black corresponds to positive/negative polarity of the LOS magnetic field. Green lines indicate PILs while red contours outline the strong Ca II H line emission. The yellow circle in panel (c) marks the center of the sheared ribbon and wedge-like structure. The white arrow in panel (b) indicates one of the transient brightenings prior to the onset of the flare.
(d-f) show the observations of SDO/HMI and AIA. The images are formatted the same as (a-c) except that the background images are filter magnetograms from HMI in units of Gauss. The yellow circle in panel (f) marks the location of the trigger region and the white arrow in panels (d, e) indicate the pre-flare brightening. The white- and black-colored broken lines in panel (f) indicate the configurations of enhanced flare ribbon.
The intensity scale saturates at $\pm 0.1$ (Stokes-V/I) in panels (a-c) and $\pm 1000$ G in panels (d-f).
}\label{fig:fig2}
\end{figure}

The lower three panels (d-f) of Figure \ref{fig:fig2} show the SDO data. The images are formatted as described for panels (a-c), but the red contours indicate the outlines of the emission in AIA 1600{\AA}. The FOV of panels (d-f) were coordinated as almost same as panels (a-c). Note that the bottom three panels show the magnetic field strength in the units of Gauss ($Mx~cm^{-2}$), while the top ones show the Stokes-V/I (dimensionless quantity). Obviously, the spatial resolution of the HMI images is lower than that of SOT, as mentioned in Section \ref{sec:data1}. Although the spatial resolution of AIA is also lower than that of SOT, the initial sheared ribbon was clearly seen in panel (f) and it will be enhanced as indicated by the broken lines. In the previous study, they could not see the very initial phase of the flare ribbons because the cadence of SOT was limited to five minutes. On the other hand, because the cadence of AIA is 45 second, here we can see the evolution process fom the pre-flare emission to the flare ribbon appearance in AIA images. We can identify again the small wedge-like structure, i.e., the trigger region in the yellow circle on panel (f). A transient brightening on the wedge-like structure is also detected as indicated by the white arrow in panels (d, e). This brightening, which is shown in panels (a, d), was first observed around 13:00 UT, i.e., about 4.5 hours prior to the flare onset. Furthermore, the location of the brightening always satisfied the condition of the RS-type configuration as seen in panels (d, e). Therefore, the same typical features of the flare trigger region observed by Hinode were observed also with SDO.

\begin{figure}
 \begin{center}
  \includegraphics[width=15cm,clip]{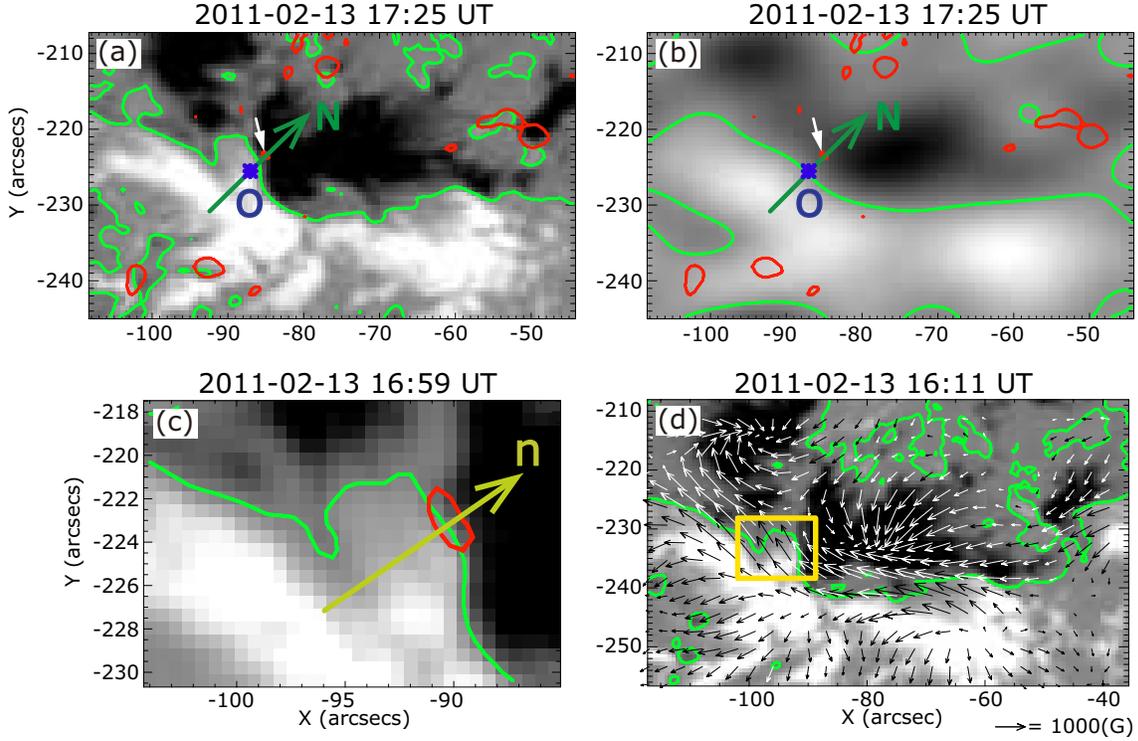} 
 \end{center}
\caption{
Images of active region and vectors $\bm{N}$ and $\bm{n}$ on the flare trigger region obtained by SDO. Panel (a) is formatted as same as Figure \ref{fig:fig2} (d-f). The background of panel (b) is the smoothed LOS magnetic field by applying low-pass filter. Green lines are smoothed PILs and point $\bm{O}$ is the point on the smoothed PIL that is nearest to the location of the last chromospheric emission before the flare onset. $\bm{N}$ is the normal vector of smoothed PIL. White arrows in panels (a, b) indicate the contour of the last chromospheric emission before the flare onset. Panel (c) is the zoom-in view of overlaid image on 16:59 UT, where vector $\bm{n}$ is orthogonal to the non-smoothed PIL of the trigger region and surrounding the chromospheric emission contour. Panel (d) shows the vector magnetograms. Grayscale indicates LOS magnetic field and overlaid arrows is transverse field. Yellow rectangle surrounds the region where the shear angle $\theta$ was calculated. The gray scale intensity is saturated at $\pm 1000$ G.
}\label{fig:fig3}
\end{figure}

We measured the angles $\theta $ and $\phi$ from the HMI images shown in Figure \ref{fig:fig3}. Panel (a) shows the filter magnetogram obtained by HMI, which is formatted as same as Figure \ref{fig:fig2} (d-f). The background of Figure \ref{fig:fig3} (b) is the smoothed LOS magnetic field shown in panel (a). We defined the point $\bm{O}$ as the nearest point to the last chromospheric emission before the flare onset (which is indicated by the white arrow in panels (a, b)) on the smoothed PIL. Then the normal vector $\bm{N}$ was defined as shown in panels (a) and (b). In panel (c), we defined the vector $\bm{n}$, which is orthogonal to the non-smoothed PIL, on the chromospheric emission region. Then we measured the angle $\phi = 336^{\circ}$-$352^{\circ}$ between vectors $\bm{N}$ and $\bm{n}$. We also measured the shear angle $\theta = 88\pm11^{\circ}$ of the vector magnetic field over the yellow rectangle in panel (d), where the error range is given by the standard deviation. Since the results using Hinode data were $\theta = 82\pm7^{\circ}$, $\phi = 318^{\circ}$-$331^{\circ}$, the angle measured by SDO are consistent with those of Hinode, and both results satisfied the condition for the RS-type flare trigger field.

\begin{figure}
 \begin{center}
  \includegraphics[width=10cm,clip]{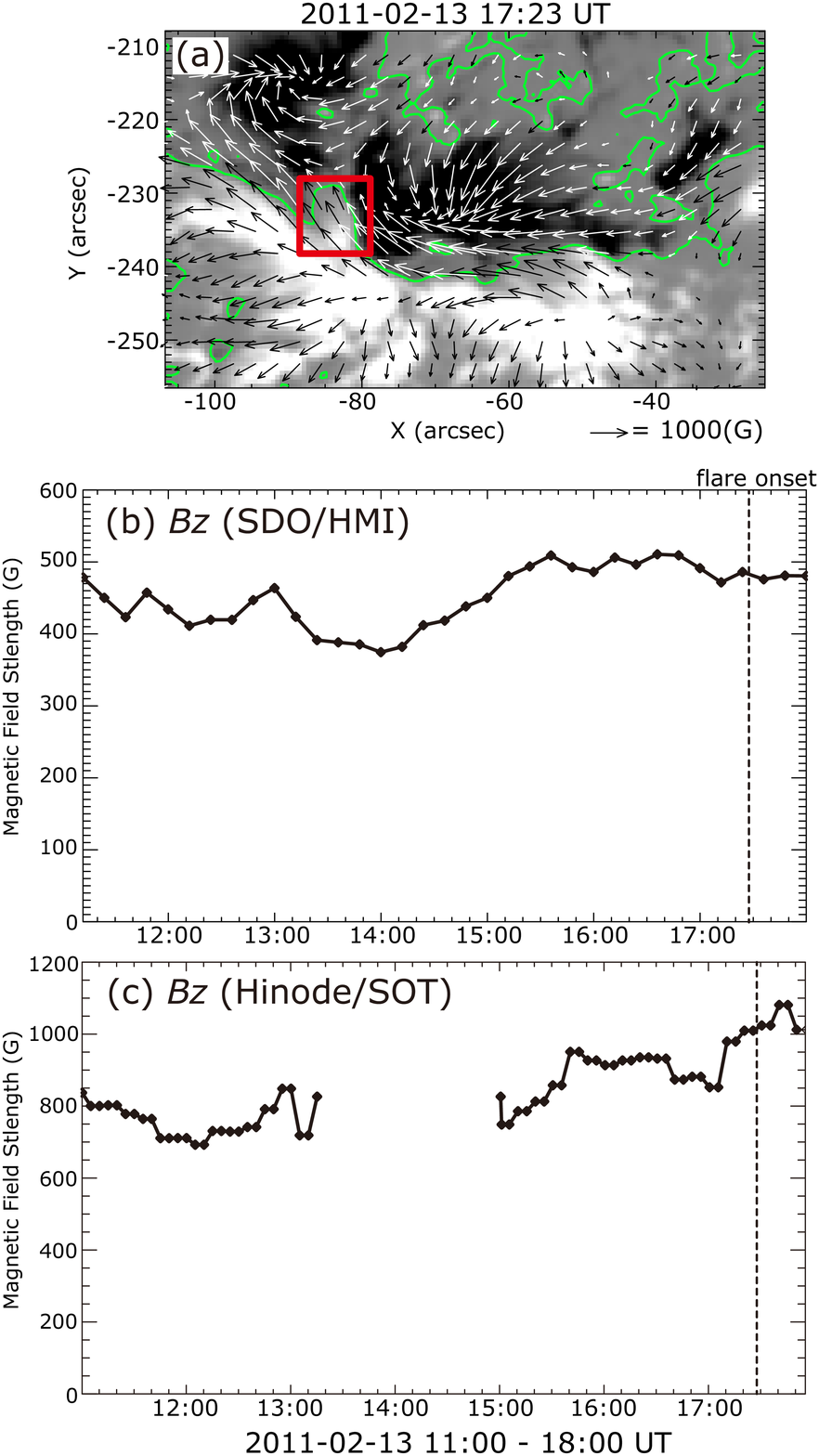} 
 \end{center}
\caption{
(a)	Vector magnetogram just prior to the onset time of the M6.6 flare obtained by SDO/HMI. The format is the same as Figure \ref{fig:fig3} (d). A red rectangle surrounds the region where the average strength of the positive magnetic field was calculated.
(b)	Temporal evolution of the average strength of the positive LOS magnetic-field $B_{z}$ observed by SDO in the trigger region. The vertical broken line indicates the onset time of the M6.6 flare. The strength of the positive magnetic field was observed at 12 minutes cadence and the strength at each time was plotted with a data point.
(c)	The temporal evolution of the average strength of the positive LOS magnetic field $B_{z}$ observed by Hinode. The format is the same as panel (b), although the cadence is five minutes.
}\label{fig:fig4}
\end{figure}

We also analyzed the temporal evolution of the average strength of the positive components of the LOS magnetic field in the trigger region surrounded by the red rectangle shown in Figure \ref{fig:fig4} (a). The rectangle encloses almost the same area as in the previous study, but the area is a little bit different from the yellow rectangle of Figure \ref{fig:fig3} (d). The temporal evolution of the magnetic field from 11:00 UT to 18:00 UT is plotted in Figure \ref{fig:fig4} (b) with the result observed by Hinode in Figure \ref{fig:fig4} (c). Here, we converted the Stokes-V/I signal to magnetic field strength in Gaussian units using the function
\begin{equation}
B_{z} = 4.30212 B_{z}^{\prime} - 33.7419,
\label{equation:gauss}
\end{equation}
which is derived from the regression line of a scatter plot of the Stokes-V/I signals and SP magnetic field scanned between 16:00-16:32 UT, where $ B_{z}^{\prime}$ and $ B_{z}$ are Stokes-V/I values and converted values (in Gaussian units), respectively. The detailed method is described by \citet{bamba13} and \citet{chae07}. The broken lines in panels (b, c) indicate the onset time of the M6.6 flare. The strength of the positive magnetic field gradually increases to the flare onset. The magnetic field strength of the trigger region calculated from the SDO data increases by 20\% from 13:30 UT to 15:40 UT, which corresponds to the time of the data gap in the SOT Stokes-V/I images. This trend is very consistent with the evolution of the average strength of the positive magnetic field calculated from the Hinode data (plotted in panel (c)), and supports the existence of magnetic inflow from the northern positive region to the triggering region reported by \citet{toriumi13}. However, the absolute value of magnetic field strength measured by Hinode and SDO do not match each other. We will discuss the quantitative differences in the magnetic field measurements in Section \ref {sec:discussion}.

As reported in subsections \ref{sec:results1} and \ref{sec:results2}, the magnetic structure observed by SDO satisfies the features of the flare trigger field proposed by \citet{kusano12}, and is consistent with the observations by Hinode. Therefore, we can conclude that the SDO data are utilizable for our flare trigger study. However, there are some discrepancies between $\theta$, $\phi$ and the magnetic field strength between the results from SDO and Hinode, and we need to be cautious interpreting them as discussed in Section \ref{sec:discussion}.

\subsection{Temporal Evolution of Angles $\theta$ and $\phi$ using SDO} \label{sec:results3}

\begin{figure}
 \begin{center}
  \includegraphics[width=15cm,clip]{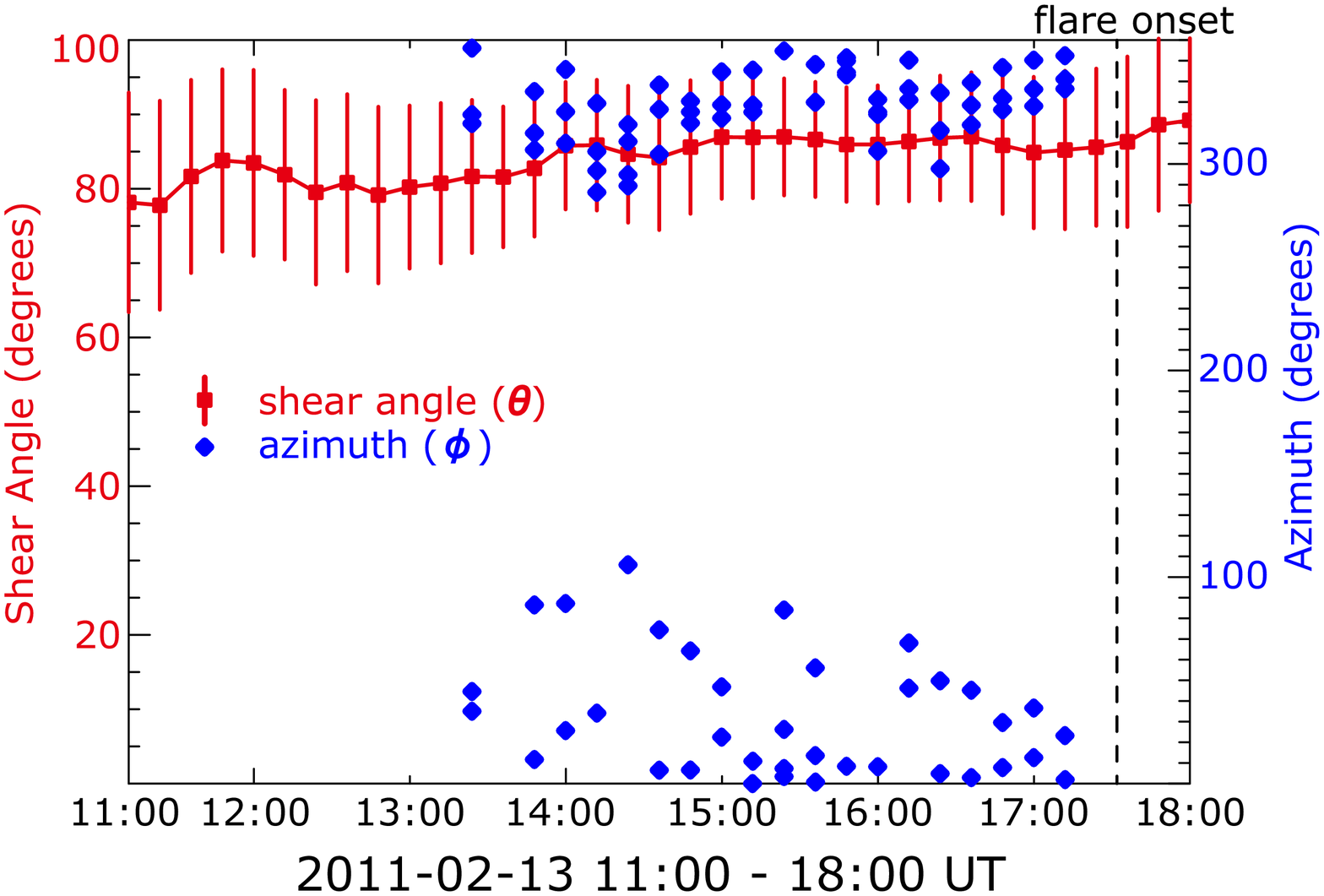}
 \end{center}
\caption{
Temporal evolution of the shear angle $\theta$ and the azimuth angle $\phi$. The values of $\theta$ from 11:00 UT to 18:00 UT were calculated with 12 minutes cadence and indicated by red-colored quadrangles. The error bar of $\theta$ is the standard deviation of multiple measurement. Blue-colored diamonds indicate the values of $\phi$ calculated from 13:23 UT to 17:11 UT. The vertical broken line indicates the onset time of M6.6 flare. The points plotted at points smaller than $100^{\circ}$ should be seen as continuously distributed to the plots above $360^{\circ}$ because we defined the angle $\phi$ in the range $0^{\circ} - 360^{\circ}$.
}\label{fig:fig5}
\end{figure}

In the previous study, since the SOT/SP measurement was taken only at 16:00-16:32 UT, they were not able to measure the shear angle $\theta$ just before the flare onset time. On the other hand, HMI had frequently taken vector magnetograms at 12 minutes cadence, and thus we can continuously observe the evolution of the $\theta$ angle from about 6.5 hours prior to the flare with HMI data. The red-colored squares plotted in Figure \ref{fig:fig5} indicate the $\theta$ angle at each time. The error bar indicates the standard deviation of the magnetic field distribution in the red rectangle in Figure \ref{fig:fig4} at each time. The result shows that the magnetic shear in the flare trigger region slightly increases until the onset time of the M6.6 flare.

Moreover, we also measured the evolution of the azimuth angle $\phi$, every 12 minutes. The measurement was taken from 13:23 UT to 17:11 UT when chromospheric emission in AIA 1600{\AA} was observed on the PIL in the trigger region. We measured the azimuth angle at five points in the pre-flare emission region at each time. The results are indicated by the blue-colored diamonds plotted in Figure \ref{fig:fig5}. Because we defined the angle $\phi$ in the range $0^{\circ} - 360^{\circ}$, the points smaller than  $\phi = 100^{\circ}$) should be seen as continuously distributed to the plots above $360^{\circ}$. The average value of $\phi$ at each time satisfied the RS-type condition, indicating that the basic structure of the flare-trigger field was already formed four hours before the flare onset.

\subsection{Temporal Evolution of Strength of LOS and Transverse Magnetic Field using SDO} \label{sec:results4}

We calculated the averaged positive field strength of the LOS magnetic field in the trigger region surrounded by red rectangle in Figure \ref{fig:fig4} (a). Here, using high-cadence vector magnetograms obtained by HMI (12 minutes cadence), we investigated the temporal evolution of the average magnetic field strength in the trigger region both for the LOS and transverse components of the magnetic field.

\begin{figure}
 \begin{center}
  \includegraphics[width=11cm,clip]{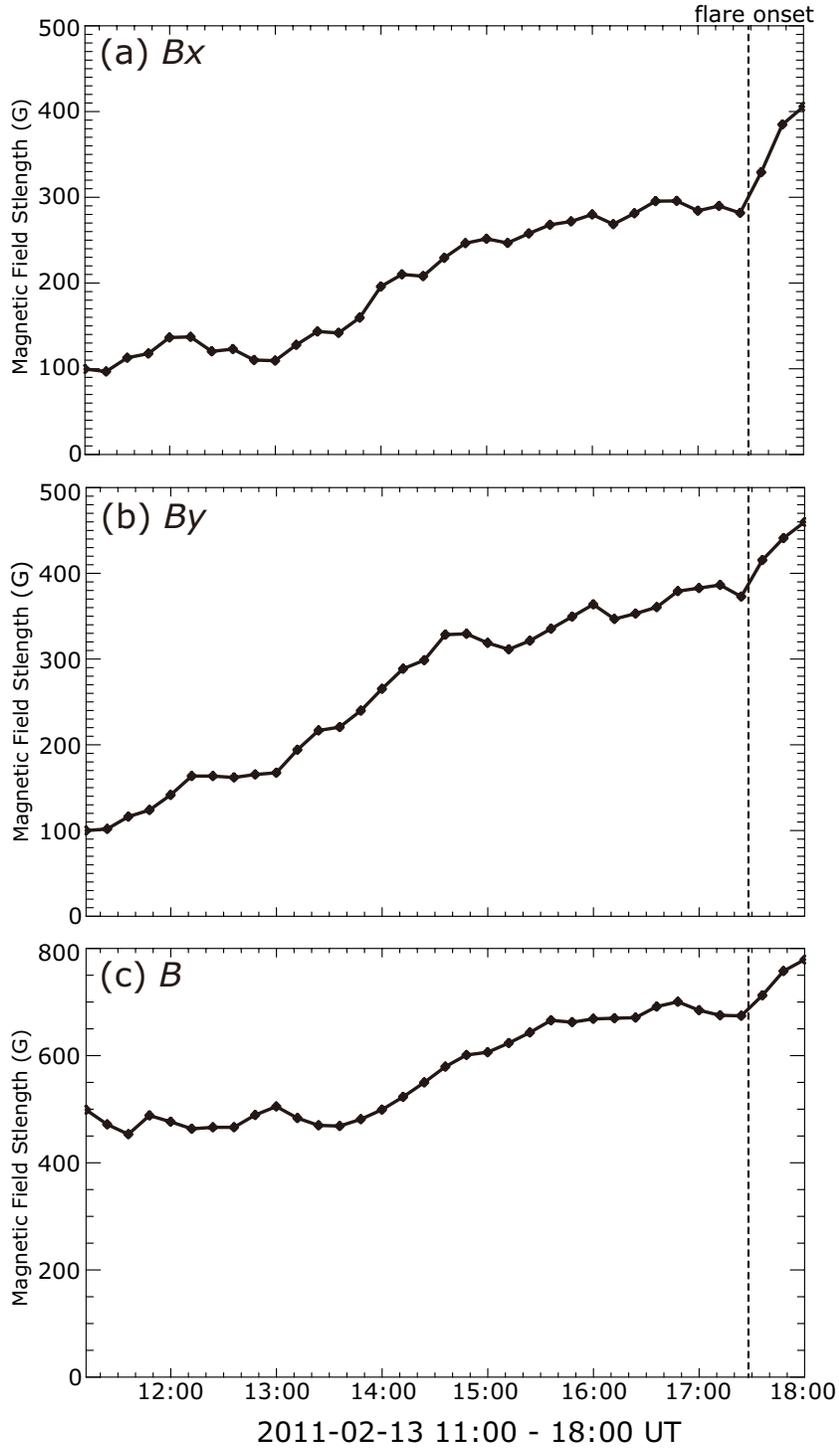} 
 \end{center}
\caption{
The temporal evolution of transverse magnetic field strength in the trigger region. The region where the magnetic field strength is calculated in the region where the LOS magnetic field is positive in red-colored rectangle in Figure \ref{fig:fig4}(a). The format of each plot is same as Figure \ref{fig:fig4}(b). Plots (a) and (b) shows the evolution of transverse field strength of $B_{x}$ (EW component) and $B_{y}$ (NS component), respectively. Plot (c) shows the evolution of $B = \sqrt{B_{x}^{2} + B_{y}^{2} + B_{z}^{2}}$.
}\label{fig:fig6}
\end{figure}

Figure \ref{fig:fig6} (a, b) shows the temporal evolution of the average magnetic field strength of the transverse components ($B_{x}$ is EW and $B_{y}$ is NS component) in the positive pole of the flare-trigger region (red square in Figure \ref{fig:fig4}). Panel (c) shows the evolution of the total magnetic field strength $B = \sqrt{B_{x}^{2} + B_{y}^{2} + B_{z}^{2}}$ in the trigger region, where $B_{z}$ is the LOS component plotted in Figure \ref{fig:fig4} (b). The strength of $B_{x}$ and $B_{y}$ increased more quickly than $B_{z}$ until the onset time of the flare, which is consistent with the magnetic shear in the trigger region being strengthened toward the flare (seen in the red-colored plot of Figure \ref{fig:fig5}). Moreover, we can see a rapid increase of magnetic field just after the flare onset. This variation is shown only in the profiles of $B_{x}$ and $B_{y}$. The immediate increase of mean horizontal field strength around the trigger region after the flare onset is consistent with the results reported in \citet{liu12}.

\section{Discussion} \label{sec:discussion}

As mentioned in Section \ref{sec:results}, we noticed that there are some discrepancies between the Hinode and SDO observations in the measurements of the angles $\theta$, $\phi$ and magnetic field strength, which are summarized in Table \ref{tab:table2}. The values of $\theta$ and $\phi$ measured with SDO were slightly larger than those measured with Hinode, although both the SDO and Hinode observations satisfied the condition of the RS-type flare-trigger field. The discrepancy in $\theta$ is not significant, and the range of $\theta$ measured by Hinode was within that of SDO. It indicates that the definition and the method of measurement of $\theta$ is robust. On the other hand, the $\phi$ angles are extremely sensitive to the shape of PIL in the trigger region and the location of point $\bm{O}$. The difference in both the resolution of the magnetic field measurements and in the chromospheric images may cause the small discrepancy in $\phi$. In addition, the slight difference of emission structure between Hinode the Ca II H line and AIA 1600{\AA} may affect the discrepancies in the chromospheric emission contours. Therefore, we should note that there is some uncertainly in our analysis method for determining the azimuth angle $\phi$ using Hinode and SDO data, and we need to be cautious in interpreting the results for the $\phi$ measurement.

\begin{table}
  \caption{Summary of the results in this study and in the previous study (\cite{bamba13}).}\label{tab:table2}
  \begin{center}
    \begin{tabular}{|c|c|c|c|}
      \hline
      & shear angle & azimuth & increment of magnetic field strength \\
      & & & (magnetic field on 11:00 UT and 18:00 UT) \\ \hline
       SDO & $88\pm11^{\circ}$ & $336^{\circ}-352^{\circ}$ & $\sim 135 ~ G ~(375 ~ G \rightarrow 510 ~ G)$ \\
   Hinode & $82\pm7^{\circ}$ & $318^{\circ}-331^{\circ}$ & $\sim 350 ~ G ~(675 ~ G \rightarrow 1025 ~ G)$ \\
      \hline
    \end{tabular}
  \end{center}
\end{table}

For the temporal evolution of the magnetic field strength, the trend that the field strength in the trigger region increased until the onset of the flare is consistent between both measurements. However, the absolute values of the magnetic field strength in the SDO measurements are smaller those from Hinode. We inferred that a difference in Figure \ref{fig:fig4}(b, c) might be due to the difference of the measurement method between SDO/HMI and Hinode/SOT. SDO obtains magnetic field by filtergraph observation while Hinode performs spectroscopic observations with SOT/SP. Therefore, we think that the values measured by Hinode are more accurate than the values measured by SDO. However, we should note that the trend of the temporal evolution of the LOS magnetic field strength is similar between Hinode and SDO, and both observations clearly showed features of the RS-type magnetic configuration. It indicates that SDO data are utilizable for the analysis of the flare trigger.

\section{Summary} \label{sec:summary}

We analyzed an M6.6 flare that occurred on 13 February 2011 with SDO data and investigated the consistency of the qualitative and quantitative features of the flare-trigger field with our previous study (\cite{bamba13}). We used HMI vector and filter magnetograms and AIA 1600{\AA} images, and we detected the trigger region with SDO data. The results of our measurement of $\theta$ and $\phi$ satisfy the RS-type condition. We also evaluated the magnetic field strength in the trigger region, and confirmed that the features of the temporal evolution of the LOS magnetic field strength calculated with SDO data were consistent with those derived from Hinode data, although the absolute value of the magnetic intensity is rather weaker than that of the Hinode observations. Therefore, we concluded that the analysis method proposed in the previous study is applicable even to SDO data. We also demonstrated that the trigger region is detectable with SDO although the spatial resolution and wavelength observed by Hinode and SDO are different. Moreover, we performed an analysis of the temporal evolution of $\theta$ \& $\phi$ and the transverse magnetic field strength with SDO data, and we found a rapid increase of the transverse magnetic field strength after the flare onset.

Based on the results above, we estimated the number of events, which are utilizable for our flare trigger study with SDO. We selected events from all flares observed by SDO on the basis of the following criteria; (1) The flare occurred on the solar disk and was located within $\pm750^{\prime\prime}$ from the disk center at least six hours before the onset time. (2) The magnitude of the flare is larger than M5.0. (3) There are continuous and steady data sets for the HMI filter and vector magnetograms and AIA 1600{\AA} images. We found that eleven X-class flares and twenty M-class flares meet our criteria between 11 February 2011 and 31 January 2014. It suggests that sufficient SDO data are available for a statistical analysis of the flare trigger process, although we have to be cautious judgement for the applicability of our method to events with more complicated magnetic field compared to the sample event. The analysis of the temporal evolution of the angles $\theta$, $\phi$ and the LOS field strength and transverse magnetic field are important to understand the critical conditions for triggering flares. The high cadence of SDO data is an advantage for such an analysis, and may be effective for revealing the flare trigger mechanism.

A statistical analysis of many flare events is required to examine the competing flare trigger models, and the high-quality data obtained by SDO is powerful for that purpose. SDO data are useful also for flare prediction experiment. On the other hand, the resolution of SDO might not be enough to track small elements forming in the flare-trigger field. Therefore, the combined analysis of Hinode and SDO will be more useful for the study of flare trigger processes.

\bigskip

The HMI and AIA data have been used courtesy of NASA/SDO and the AIA and HMI science teams. Hinode is a Japanese mission developed and launched by ISAS/JAXA, which collaborates with NAOJ as a domestic partner and with NASA and STFC (UK) as international partners. Scientific operation of the Hinode mission is conducted by the Hinode science team organized at ISAS/JAXA. This team mainly consists of scientists from institutes in the partner countries. Support for the post-launch operation is provided by JAXA and NAOJ (Japan), STFC (UK), NASA, ESA, and NSC (Norway). This work was partly carried out at the NAOJ Hinode Science Center, which is supported by the Grants-in-Aid for Creative Scientific Research ``The Basic Study of Space Weather Prediction" (Head Investigator: K. Shibata) from the Ministry of Education, Culture Sports, Science and Technology (MEXT), Japan, by generous donations from Sun Microsystems, and by NAOJ internal funding. Part of this work was carried out on the Solar Data Analysis System operated by the Astronomy Data Center in cooperation with the Hinode Science Center of the NAOJ. This work was supported by the Grants-in-Aid for Creative Scientific Research (B) ``Understanding and Prediction of Triggering Solar Flares'' (head investigator: K. Kusano), from the Ministry of Education, Culture, Sports, Science, and technology (MEXT), Japan. This work was also supported by JSPS KAKENHI Grant Number 23540278 and JSPS Core-to-Core Program 22001.


\end{document}